\documentclass[twocolumn,showpacs]{revtex4}

\usepackage{epsf}
\usepackage{graphicx}
\usepackage{dcolumn}
\usepackage{amssymb}

\newcommand{\BE}{\begin{equation}}
\newcommand{\EE}{\end{equation}}
\newcommand{\BA}{\begin{eqnarray}}
\newcommand{\EA}{\end{eqnarray}}

\def\be{\begin{equation}}
\def\ee{\end{equation}}
\def\bea{\begin{eqnarray}}
\def\eea{\end{eqnarray}}

\begin{document}
\input epsf
\renewcommand{\topfraction}{0.8}

\title {\Large\bf A new two-faced scalar solution and cosmological SUSY breaking.}

\author{\bf Marina Shmakova}
\affiliation{KIPAC, Stanford, CA 94025, USA}

\author{\bf Valentin Burov}
\affiliation{Physics Department, Moscow State University, Moscow,
Russia}

{\begin{abstract} We propose a possible new way to resolve
the long standing problem of strong supersymmetry breaking coexisting
 with a small cosmological constant. We consider a scalar component of a minimally coupled N=1 supermultiplet
in a general Friedmann-Robertson-Walker (FRW) expanding universe.
We argue that a tiny term, proportional to  $H^2\sim 10^{-122}$ in Plank's units, appearing in the field equations
due to this expansion will provide both, the small vacuum energy and the heavy mass of the scalar supersymmetric
partner. We present a non-perturbative solution for the scalar field with an unusual
 dual-frequency behavior. This solution has two characteristic mass scales related to the Hubble parameter as $H^{1/4}$ and $H^{1/2}$ measured in Plank's units.
\end{abstract}}
\maketitle

\section{Introduction}



A discrepancy between the vacuum
energy $\sim M_{P}^4$ predicted by quantum field theory and the observed tiny cosmological constant
$\Lambda \sim 10^{-122}M_{P}^4 $ is one of the major problems
in reconciling quantum field theory with cosmology.
Supersymmetry could partially resolve
this problem because the contributions to the vacuum energy from
fermionic and bosonic loops cancel each other giving zero vacuum
energy and a zero cosmological constant.  The unsatisfactory feature of
this scenario is the fact that supersymmetry should be  broken at a rather
large scale (compared to the cosmological constant) to give the
observed shifts in the masses of superpartners.

We propose a possible new way to resolve
this long standing problem of strong supersymmetry breaking coexisting
 with a small cosmological constant.
We consider a scalar component of minimally coupled N=1 supermultiplet
in a general Friedmann-Robertson-Walker (FRW) expanding universe.
We argue that a tiny term, proportional to  $H^2\sim 10^{-122}$ in Plank's units, appearing in the field equations
due to this expansion will influence
 both the vacuum energy and the masses of supersymmetric
partners in a very surprising way.
We were able to find an unusual
 non-perturbative solution for the scalar field with the double periodic behavior of non-linear oscillator.
 This "two-faced" behavior provides a singular scalar solution with two very different frequencies  corresponding to
mass scales of  $\sim 2 \cdot 10^{-3}$eV and $\sim 300$GeV. It is important to emphasize that these two mass scales are not related to regular spontaneous symmetry breaking mechanism.

A straightforward
estimation  shows that the
size of the universe changes during the time of the
virtual loop fluctuations, $t_{loop} \sim t_P$ where $t_P$ is the Plank time ($ t_P \sim 10^{-43} s$).
The Plank scale is the natural choice because it represents the most
energetic part of the fluctuation spectra.
During this time the
 relative change of the size of the universe $R_0$ will be proportional to $ \frac{c t_P}{R_0} \sim 10^{-61}.$
 The change in the scale factor $a$  during this time can be
 estimated using the value of the Hubble parameter $H = \frac{\dot a }{a}.$
 For observed $ H \sim 70$ {\small km/s/Mpc} $\sim 2.3\cdot
10^{-18} s^{-1}$ the change in the scale factor
 $\Delta (a^2) = (\dot a t_P)^2$ will be proportional to:
 $$ \frac{\Delta (a^2)}{a^2}=H^2 (t_P)^2 \sim 6 \cdot 10^{-122}.$$


 This means that the universe will change over the characteristic virtual loop time (Plank time $t_P$)
and that the interactions of virtual particles in vacuum loops should reflect this difference
This change in the virtual particles at the beginning and at the end of a vacuum loop will lead to the effective "decompensation"
on a scale of $\sim  10^{-122}.$
This amazing coincidence between tiny value of $\frac{\Delta (a^2)}{a^2}$  and the ratio of  ${\Delta E_{vac}\over M_{Plank}^4}$
may be related to the statement made
 by P. Dirac  about the "non-random" nature of the coincidence of
 the large dimensionless values in physics \cite{Dirac:1975vq}.


 The regular SUSY models in a flat "static" space have a zero
 vacuum energy due to the cancelation of bosonic and fermionic contributions \cite{Martin:1997ns}.
  However, the expansion of the universe could break this balance and create a non-zero vacuum energy density.
 In the expanding universe the fermionic and bosonic  vacuum fluctuations behave differently
 and the effects on the energy density will be of the order of $\rho \sim 10^{-122} M_P^4.$
The estimations of the scalar component contribution to the energy density fluctuation in the expanding universe can be found in multiple papers:
\cite{Linde:1982,Howking:1983,Allen:1985ux}. Recently, similar result for supersymmetric model was derived in \cite{Bilic:2010}. As it follows from these papers the scalar energy density is $\rho_s \sim H^2$ with $M_\phi  \approx  10^{-30} M_P \approx 3 \cdot 10^{-3} eV.$ The fermionic contribution was considered in \cite{Allen:1986qj}.
 Following this paper the fermionic contribution is proportional  to $\rho_f \sim R_0^{-3}$ where $R_0$ is a horizon size. For the current epoch this
 estimation gives a very small value $\rho_f  \approx (10^{-22})^4 eV^4 $.

In this paper we will investigate the possibility of a novel approach to supersymmetry  breaking
that will provide relatively high fermion-scalar mass splitting ($10^2 - 10^3$ GeV) while keeping the
value of the vacuum energy density less than $(10^{-2})^4 eV^4$.
  We propose an alternative approach based on a non-perturbative solution for scalar field in the expanding Friedmann-Robertson-Walker background.

\section{Non-perturbative scalar solution}

General N=1 supersymmetric Lagrangian of  chiral
supermultiplets (see \cite{Martin:1997ns} and references therein) has a form: \bea L &=& \partial_\mu \phi_i^\dagger
\partial^\mu \phi_i - \vert m_{ij}\phi_j + \lambda_{ijk} \phi_j
\phi_k \vert^2 \\
&+& \frac{\rm i}{2} \bar \psi_i \gamma^\mu
\partial_\mu \psi_i - \frac{\rm i}{2}m_{ij}\bar\psi_i \psi_j \nonumber \\
&-& \frac{\lambda_{ijk}}{2} ( \bar\psi_i \psi_j -\bar\psi_i \gamma_5
\psi_j ) \phi_k -\frac{\lambda_{ijk}}{2} ( \bar\psi_i \psi_j +
\bar\psi_i \gamma_5 \psi_j )\phi_k , \nonumber
\eea
where $\phi _i$ is a set of scalar components of chiral multiplets
and $\psi_i$ are Majorana spinors from the same multiplets.
We will concentrate on a single massless  scalar field with a quartic self-interaction
in a simplest form:
$$
 L = \partial^\mu \phi \partial^\mu \phi - \lambda \phi^4 .
$$
For the expanding universe with a Hubble parameter $H =\frac{\dot a}{a}$
 the equation of motion for the field $\phi $ has a form:
$$ \ddot \phi + 3  H \dot \phi + \frac{\partial V}{\partial \phi}= 0.$$
With the variable transformation:
$$
\phi (t) = \frac{u(t)}{a(t)^{3/2}},
$$ the equation of motion becomes:
\bea
\ddot u - \frac{3}{2}\left( \frac{\ddot a }{a}+\frac{1}{2}\left(\frac{\dot a }{a} \right)^2 \right) u + a^{3/2} V^\prime (u/a^{3/2})=0,
\label{eqqforu}
\eea
where an overdot indicates a derivative with respect to $t$ and a prime is a derivative with respect to the field $\phi.$ The second, "effective mass" linear term: $ -\frac{3}{2}\left( \frac{1}{2}H^2 + \frac{\ddot a }{a} \right) $
 can be estimated using Friedmann-Robertson-Walker equations \cite{Carroll:2000fy,Trodden:2004st} with the metric $ d s^2 = -d t^2 +a^2(t) d x^2 $
$$ \frac{\ddot a}{a} = - \frac{4 \pi}{3}G(\rho+3p); \,\,\,\,\, H^2=\frac{\dot a^2}{a^2} = \frac{8 \pi}{3}G \rho.
$$
The linear "mass" term related to the deceleration parameter $q=- \frac{\ddot a }{a H^2}$. The current value of the deceleration parameter is $q_0 \sim - 0.5$ and the mass term can be estimated as $ -\frac{3}{2}\left( \frac{1}{2}H^2 + \frac{\ddot a }{a} \right) \approx -1.5 \cdot H^2 \approx - H^2.$ In the early universe the value for the deceleration  parameter was probably quite different. In the case of hot relativistic gas dominance, $p=\rho/3,$ and the  mass term will flip sign as $ -\frac{3}{2}\left( \frac{1}{2}H^2 + \frac{\ddot a }{a} \right) \approx  \frac{9}{4} H^2$. We will concentrate on the present moment, in this case the approximate equation will take form:
\bea\ddot u - H^2 u + \frac{\lambda}{a^3} u^3=0\label{eqq}, \eea
for simplicity we set the numerical factor $\sim O(1)$ in front of the second term to $1.$
This non-linear oscillatory equation has two non-perturbative solutions:
\bea u_{1,2}=\pm i \frac{\sqrt{-c_1}}{\kappa } \, {\rm sn}\left[\kappa (t-c_2) ;-1-\frac{H^2}{\kappa^2} \right],
\label{eqqsol}
\eea
where ${\rm sn}\left[\kappa t ;-1-\frac{H^2}{\kappa^2} \right] $ is the  Jacobi elliptic function \cite{Whittaker}, parameter $\kappa$ is
\bea \kappa =\sqrt{\frac{\lambda c_1}{a^3 H^2 + \sqrt{a^6 H^4 + 2 a^3 \lambda c_1}}},
\label{eqqsolkk}
\eea
and $c_1$ and $c_2$ are free parameters. The parameter $c_2$ is an arbitrary time-shift that we will set to zero in all further expressions.
If the mass term changes sign, the equation
$$\ddot u + H^2 u + \frac{\lambda}{a^3} u^3=0.$$
will have solutions of the form:
$$\tilde u_{1,2}=\mp i \frac{\sqrt{-c_1}}{\tilde \kappa} \,{\rm sn}\left[\tilde \kappa t ;-1+\frac{H^2}{\tilde \kappa^2} \right],
$$
where  $\tilde \kappa=\sqrt{-\frac{\lambda c_1}{a^3 H^2 + \sqrt{a^6 H^4 + 2 a^3 \lambda c_1}}}.$

We set the value of the self-interaction constant to $\lambda \approx 10^{-12},$  the choice of the stronger self-interaction constant $\lambda \approx 10^{-2}$  is also possible, see \cite{Linde:2005ht}. This specific value of $\lambda$ will affect only the value of the free parameter $c_1$ and is not crucial for the general results of this paper.  The current value of the scale factor $a$ can be set to  $a=1,$ other possible choices discussed in \cite{Trodden:2004st}. The value of parameter $c_1$ will be  considered later.

 The Jacobi elliptic function ${\rm sn}[\tau,k]$  is one of the 12 Jacobi functions of this kind, \cite{Whittaker}. The second argument,
  $k$ is the elliptical modulus related to parameter $m$ as $k^2=m$.  In our solution $m =-1-\frac{H^2}{\kappa^2}$ with $\kappa$ defined in eq. (\ref{eqqsolkk}).
The presence of a tiny second term in $m$ proportional to $H^2$ will lead to very important consequences discussed later.
The main feature of the Jacobi elliptic functions is the existence of {\sl two distinct complex non-collinear oscillatory periods}. The two periods of elliptic ${\rm sn}[\tau,m ]$  are defined as follows
$$
{\rm sn}[\tau + 4 p K+ i 2 n K^\prime , m]={\rm sn}[\tau, m],
$$
where $p$ and $n$ are integer numbers,  and $K(m)$ is the complete elliptic integral of the first kind with $K^\prime(m)=K(1-m).$  The real physical time $t$ is related to a dimensionless complex parameter $\tau$ as $\tau =\kappa t$. It means that the direction of $\tau $ evolution with physical time $t$ on a complex plane will be defined by $\kappa. $

The non-linear equation of the type (\ref{eqq}) for particle physics was previously studied by many authors, including  M.~Fraska in \cite{Frasca:All}. In these papers, however, the second term of equation (\ref{eqq}) proportional to $H^2$ was missing and the value of parameter $m=-1.$
\section{Double-periodic solutions}

 The double-periodic nature of solution (\ref{eqqsol}) with $H^2$ dependence opens a very interesting possibility of existence of two distinct non-linear oscillatory processes in  a single two-faced solution for the scalar field $\phi$ with drastically different effective frequencies. We will show that the heavy component can have an effective mass of the order $\sim 300 - 400 $ GeV and the light component's mass can be of the order of $\sim (2 - 3)\cdot 10^{-3}$eV for the current epoch. These numbers will depend on the cosmological time and the corresponding value of the Hubble parameter. Interestingly, this dependence is very soft and smooth for the time period between the end of inflation and current time. During the inflation and reheating both scales will follow and, possibly, influence the metric evolution.

\begin{figure}[h!]
\centering {\leavevmode 
\includegraphics[width=8.5cm,height=5.5cm]{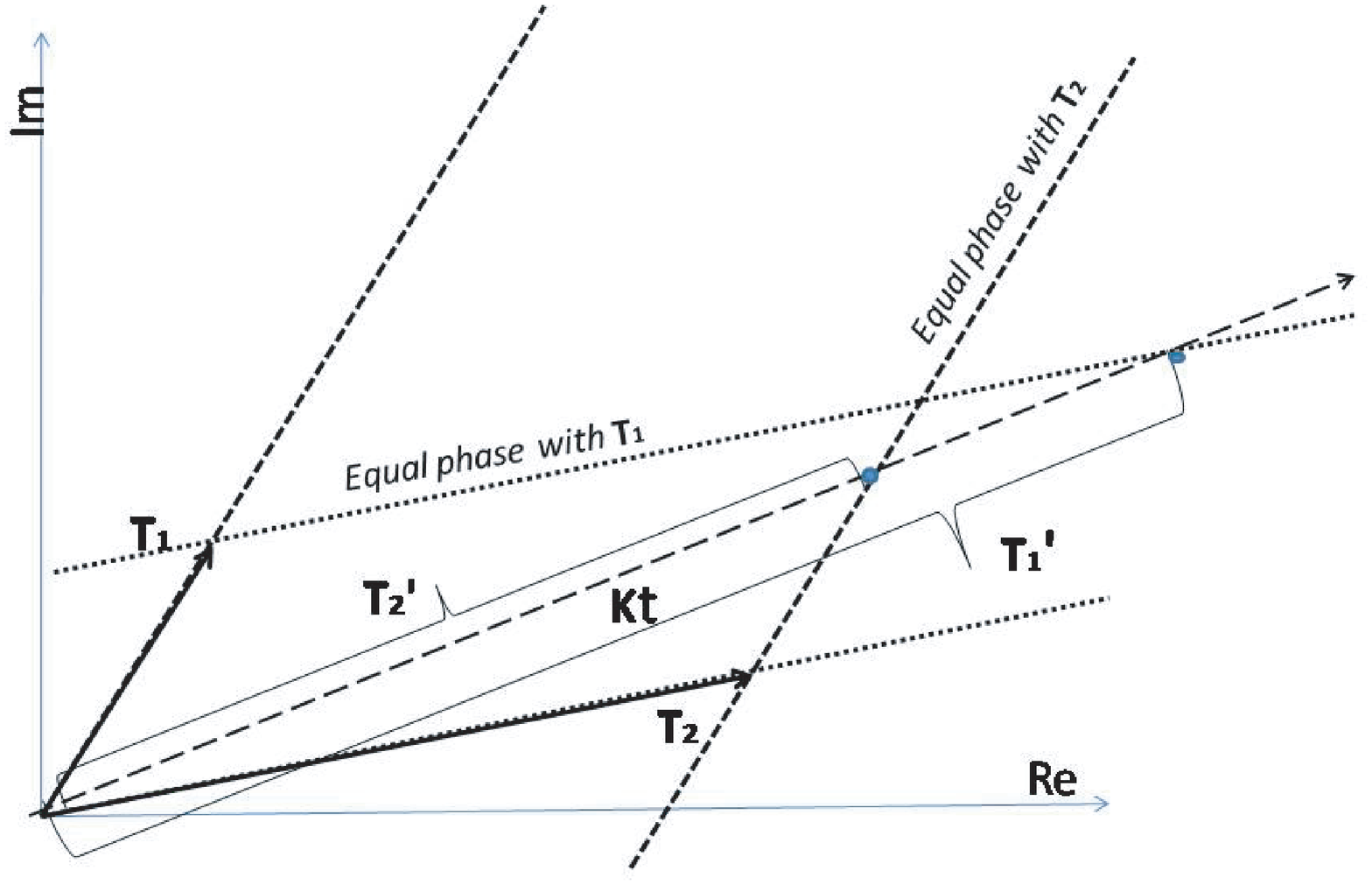}}
\caption[periods] {\centering The periods $T_1^\prime $ and $T_2^\prime $ define the physical periods along $\kappa t $ direction}.\label{periods}
\end{figure}

The periods  $T_1 = \frac{4K(m)}{|\kappa|}$ and $T_2 = \frac{i2K(1-m)}{|\kappa|}$   are complex valued and have different directions on a complex plane. The real physical periods $T_1\prime $ and $T_2\prime $ and the corresponding frequencies of the non-linear oscillator (\ref{eqqsol}) are not directly defined by values of $T_1$ and $T_2$ but are related to them through the parallelogram projections on a complex plane (see Figure 1). With the evolution of the real physical time the dimensionless argument $\tau $ of the Jacobi $\rm sn [\tau, m]$
function evolves in the direction of $\kappa.$

\begin{figure}[h!]
\centering {\leavevmode 
\includegraphics[width=8cm,height=6.5cm]{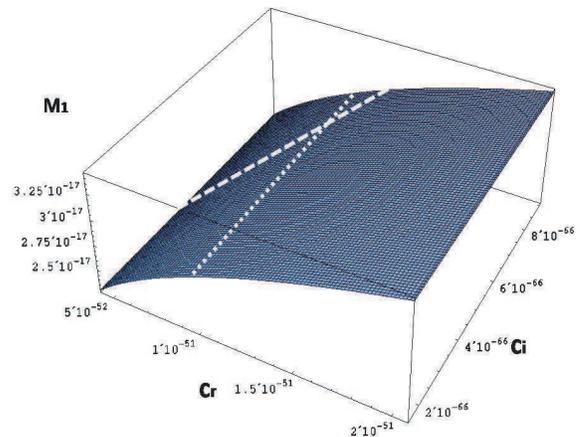}}
\caption[mass1]{\centering The heavy $M_1$ mass component of the scalar field eq.(\ref{eqqsol}) for the Hubble value $H=10^{-61} t_P$. The dotted line corresponds to the set of $c_r$ and $c_i$ values for $M_1 \approx 3 \,\cdot 10^{2}$GeV and the dashed line indicates $c_r$ and $c_i$ region for the light mass $M_2 \approx 2.8 \cdot 10^{-3}$eV.}
\end{figure}

\begin{figure}[h!]
\centering {\leavevmode 
\includegraphics[width=8cm,height=6.5cm]{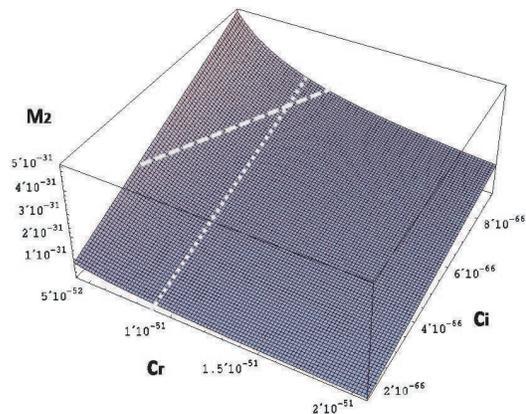}}
\caption[mass2]{\centering The light $M_2$ mass of the scalar field eq.(\ref{eqqsol}) for the Hubble parameter $H=10^{-61} t_P.$ The dashed line indicates $c_r$ and $c_i$ region for the light mass $M_2 \approx 2.8 \cdot 10^{-3}$eV and the dotted line corresponds to the set of $c_r$ and $c_i$ values for $M_1 \approx 3 \,\cdot 10^{2}$GeV.}
\end{figure}

The periodic structure of Jacobi ${\rm sn}$  function is plotted in figure 1  as a periodic set of parallelograms with side lengths $T_1$ and $T_2.$
The physical periods $T_1^\prime $ and $T_2^\prime $ indicated in figure 1 are defined as the distances between crossing points of the line in the $\kappa t $ direction with the lines of the periodic structure. These periods can  be expressed through values of $T_1,\,T_2$ and $\kappa$ as follows:
 \bea
 T_1^\prime &=& |\kappa | \frac{{\rm Im}T_2 \,{\rm Re}T_1 - {\rm Re}T_2 \,{\rm Im}T_1 }{{\rm Im}\kappa \,{\rm Re}T_1 - {\rm Re}\kappa \,{\rm Im}T_1} ; \label{eqqperiods}\\
 T_1^\prime &=& |\kappa | \frac{{\rm Im}T_1 \,{\rm Re}T_2 - {\rm Re}T_1 \,{\rm Im}T_2 }{{\rm Im}\kappa \,{\rm Re}T_2 - {\rm Re}\kappa \,{\rm Im}T_2}.  \nonumber
 \eea

The geometry of periodic structure shows that even for comparable values of $T_1$ and $T_2$ periods the physical periods $T_1^\prime $ and $T_2^\prime $ of the non-linear oscillator eq. (\ref{eqqsol}) can differ by ten or more orders of magnitude depending on the relative directions of $\kappa,\, T_1,$ and $T_2$.

The values $T_1$ and $T_2$ will depend on Hubble parameter $H$, self-interaction constant $\lambda$, scale factor $a$,  and  complex valued constant $c_1 = c_r + i \, c_i $. All values will be expressed in Plank units. The value for parameter $c_1$ can be estimated from the mass of the fluctuating field $M_\phi \approx 3 \cdot 10^{-3} eV$ related to the current value of  the cosmological constant. Figures 2 and 3 plot the behavior of the light and heavy scalar masses for the current value of the  Hubble parameter $H=10^{-61} t_P^{-1}.$ It is clear from these figures that the heavy mass $M_1 \approx 3 \,\cdot 10^{2}$ GeV and the light mass  $M_2 \approx 2 - 3 \,\cdot 10^{-3}$eV can be found simultaneously for some value of parameter $c_1.$ The real component $c_r$ of this parameter will mostly influence the heavy mass and its imaginary part $c_i$ will mostly affect the light one. We set the values of the constants $c_r$ and $c_i$ based on the known value of the cosmological vacuum energy estimated from $H^2$ and the wide predicted range of supersymmetry breaking $\sim 10^2 - 10^3$ GeV. Similar plots were also done for $H=10^{-40} t_P^{-1}$ with analogous results.  A fundamental and clear physical and mathematical criteria for the choice of the constant $c_1$ is yet to be found.

The direct calculation of the time dependence of the Jacobi sn function for the current value of Hubble parameter $H=10^{-61} t_P^{-1}$  as well as for the $H=10^{-40} t_P^{-1}$ for the first milliseconds of the universe exhibit the periodic behavior of the solution eq. (\ref{eqqsol}) with the periods (\ref{eqqperiods}).
Figures 4 and 5 demonstrate  this dual-periodic behavior for the low and high frequency oscillations for the specific choice of the parameter $c_1$. The high frequency mode is very close to harmonic oscillations (although the second derivative clearly shows the deviations from the harmonic behavior).  It is extremely hard to plot the low frequency processes with a background of a higher amplitude  and a higher frequency (up to 10-16 orders of magnitude) oscillations. We used moving average filtration to remove the high frequency oscillations. The highly non-harmonic, periodical pulse process for $H=10^{-40} t_P^{-1}$ is shown in the figure 5.

\begin{figure}[h!]
\centering {\leavevmode 
\includegraphics[width=4.2cm,height=3.2cm]{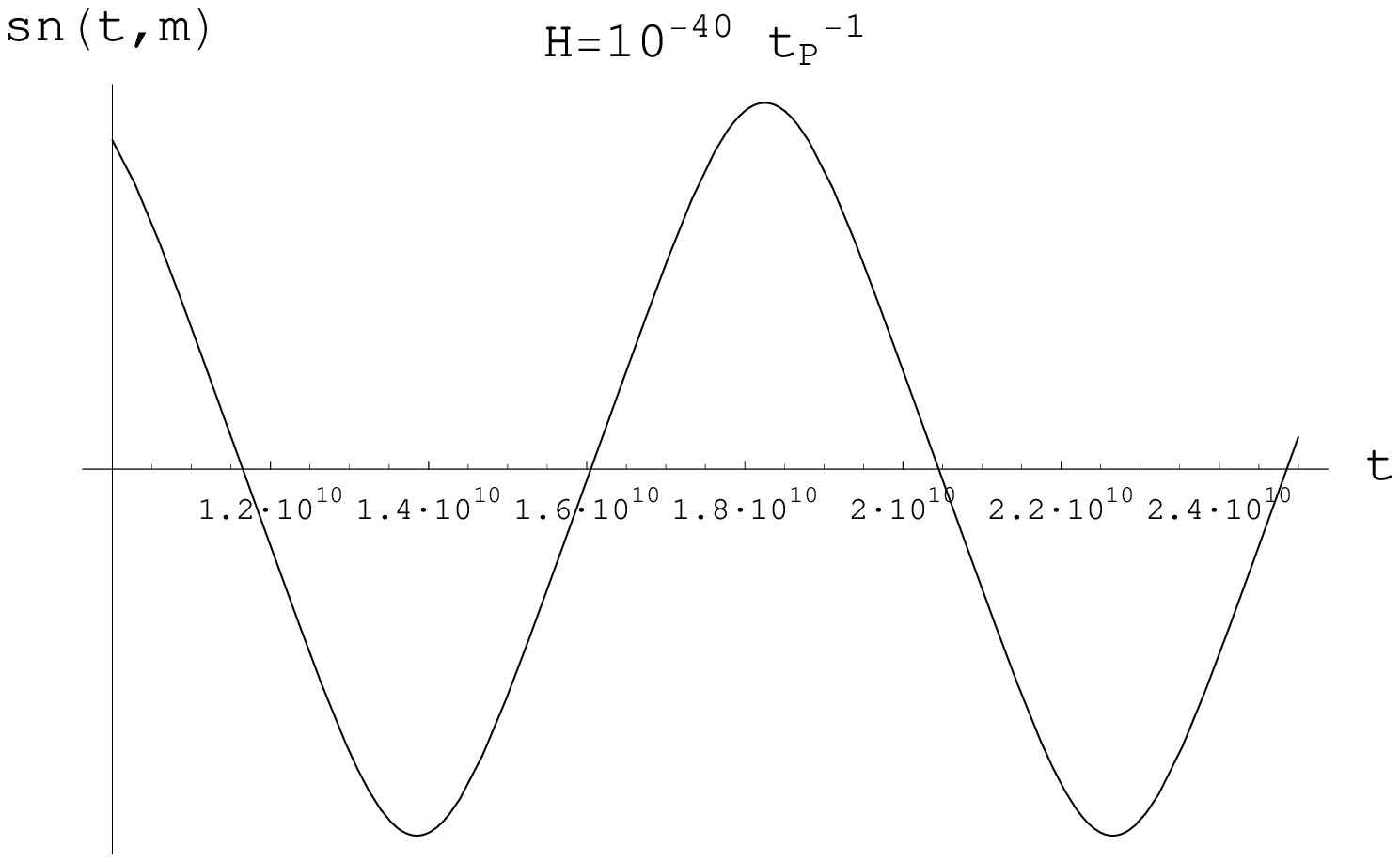}\,
\includegraphics[width=4.2cm,height=3.2cm]{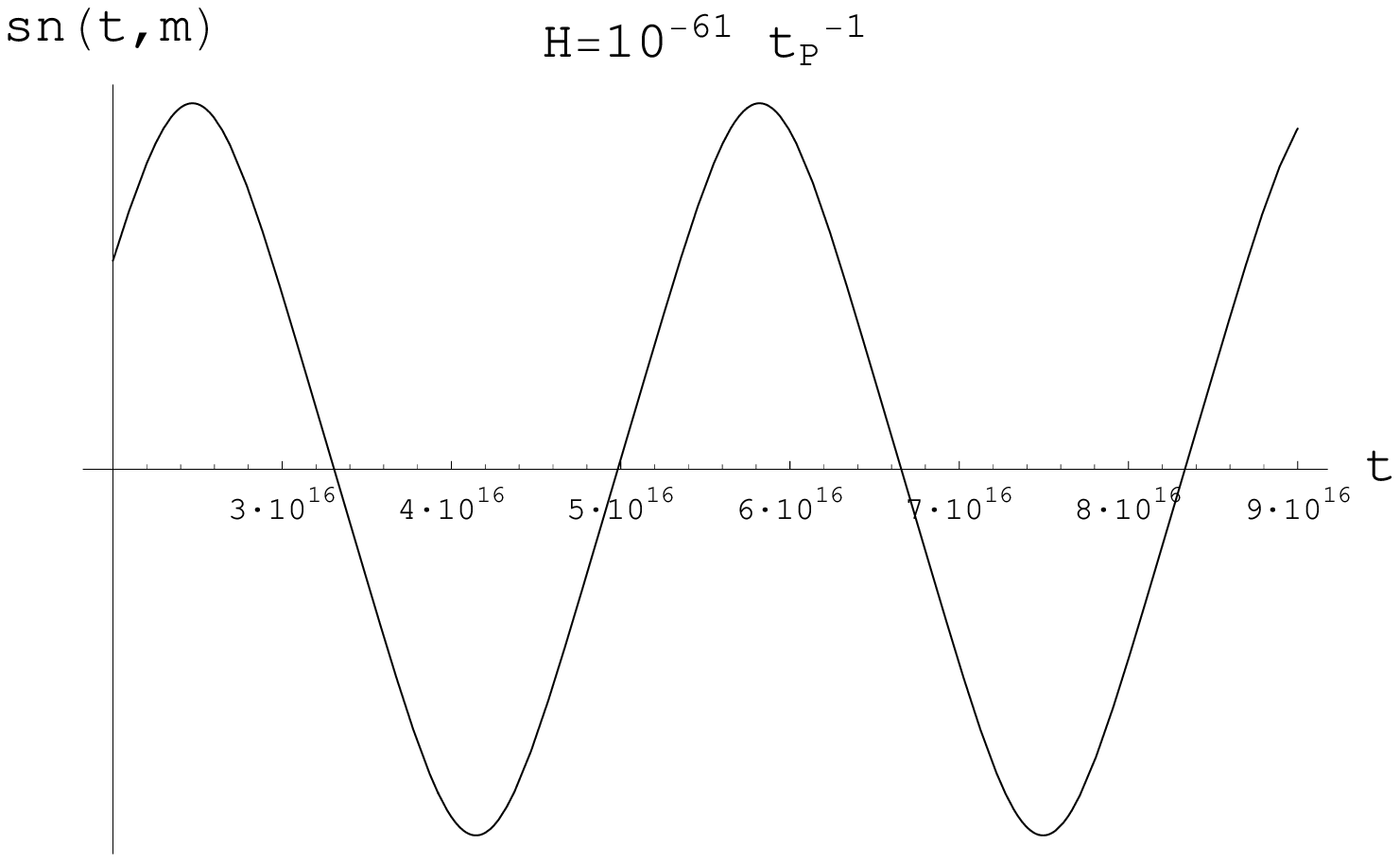}}
\caption[periods] {\centering The picture presents the high frequency, heavy mass, component of the scalar field (\ref{eqqsol}).
The left plot is for the Hubble parameter $H=10^{-40} t_P^{-1}.$ For this case the period is $T_1^\prime \approx 10^{10} t_P$
and the corresponding mass component is $M_{1}= 10^{-10} M_P$  $\approx 10^9$Gev . The right plot is for the Hubble value $H=10^{-61} t_P^{-1}$ with the corresponding period  $T_1^\prime = 3.2 \cdot 10^{16} t_P$ and corresponding mass  $M_{1}\approx 3 \cdot 10^{-17}M_P \approx 300$ Gev .}
\end{figure}

\begin{figure}[h!]
\centering {\leavevmode 
\includegraphics[width=8.5cm,height=6cm]{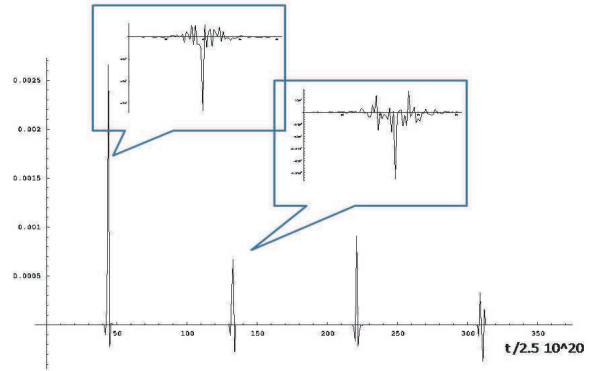}}
\caption[periods] {\centering The picture presents the low frequency, light mass, component of the scalar field (\ref{eqqsol}) for $H=10^{-40} t_P^{-1}.$
For this case the period is $T_2^\prime \approx 2\cdot 10^{22} t_P$ and the corresponding mass component is $M_{2}= 5 \cdot 10^{-21} M_P$  $\approx 50$ MeV. The low frequency process is highly non-harmonic.}
\end{figure}

\section{Conclusion}
 We were able to find a non-perturbative solution for the scalar field with an unusual
 two-faced behavior. This solution has two characteristic mass scales related to the Hubble parameter as $H^{1/4}$ and $H^{1/2}$.
  Even the possibility of a dual-frequency solution with one high mass value for a scalar superpartner of N=1 SUSY multiplet is itself a strong breaking of supersymmetry. It is important to notice that these two mass scales are not related to the usual energy levels of quantum oscillators and represent the fundamental, inner property of the scalar solution (\ref{eqqsol}). Possibly, in some cases this solution could be regarded as strongly coupled oscillators with two different resonant frequencies. Interestingly, the existence of this solution with its high and low masses is based on the appearance of a tiny term proportional to $H^2$ in the wave equation. This term is related to the third natural scale with masses $\sim 10^{-33}$eV connected to the inner frequencies of the entire universe. The possibility of such scalars with the masses quantized in terms of cosmological constant $\Lambda=3H^2 $ as $\frac{m^2}{H^2}=n $ in the framework of extended supergravity in de Sitter vacua was discussed in \cite{Kallosh:2001gr,Kallosh:2002wj}.

The results presented above leave a few unresolved questions. It is important to notice that the derivation of the above results was based on  a quasi-classical rather than  quantum approach. Our consideration of a static case allows us to replace the non-linear field equation with the non-linear oscillatory equation.
One of the major questions is the observational absence of the light scalar particles with the masses $2.5 \cdot 10^{-3}$eV. Apart from the obvious observational difficulties detecting such particles there exist additional circumstances that could make the detection especially difficult.  This light scalar will interact with its own fluctuations with the same energies as the mass of the particle. In this case an additional non-linear dumping will inevitably appear, similar to the non-linear attenuation of the waves in a diffuse noisy field. Therefore, the life time of this light particle can be very short, making it hard to detect.

Another question is related to the quantum fluctuations of the high frequency  scalar component.
 Clearly, a truly exhaustive answer can be obtained only in a framework of a supersymmetric quantum field theory in expanding cosmology.

The solution described in this paper, where the scalar superpartner has such an unusual structure is considerably novel. If there is a similar structure for the
vector multiplets is also not clear and can be a subject of future research.

The processes in the early universe during the inflation are related to fast changes of Hubble parameter as well as the scale factor $a$, and its derivatives $\dot a$ and $\ddot a$  \cite{Linde:2005ht}. The linear term in eq.(\ref{eqq}) was relatively large at this time. Additionally, the sign of this term will depend on $\frac{\ddot a}{a}$. This creates a potential feedback that can induce stabilizing/destabilizing effects on the expansion of universe.

Work of M. Shmakova was supported in part by the U.S. Department of Energy under contract number DE-AC02-76SF00515.


\end{document}